\documentclass[useAMS,usenatbib]{mn2e}
\usepackage{graphicx}
\usepackage[T1]{fontenc}
\usepackage{aecompl}
\title[From the ``blazar sequence'' to unification of blazars and radio galaxies]{From the ``blazar sequence'' to unification of blazars and radio galaxies}
\author[Dingrong Xiong, Xiong Zhang, Jinming Bai and Haojing Zhang]{Dingrong Xiong$^{1,2}$, Xiong Zhang$^{3}\thanks{E-mail: ynzx@yeah.net}$, Jinming Bai$^{1}$ and Haojing Zhang$^{3}$\\
$^{1}$National Astronomical Observatories/Yunnan Observatories, Chinese Academy of Sciences, Kunming 650011, China\\
$^{2}$The Graduate School of Chinese Academy of Sciences, Beijing
100049, China\\
$^{3}$Department of Physics, Yunnan Normal University, Kunming
650500, China}
\begin{document}

\pagerange{\pageref{firstpage}--\pageref{lastpage}} \pubyear{2002}

\maketitle

\label{firstpage}

\begin{abstract}
Based on a large Fermi blazar sample, the blazar sequence
(synchrotron peak frequency $\nu_{\rm peak}$ versus synchrotron peak
luminosity $L_{\rm peak}$) is revisited. It is found that there is
significant anti-correlation between $\nu_{\rm peak}$ and $L_{\rm
peak}$ for blazars. However, after being Doppler corrected, the
anti-correlation disappears. The jet cavity power ($P_{\rm jet}$) is
estimated from extended radio luminosity. So it is free of beaming
effect. We find that there are significant anti-correlations between
$P_{\rm jet}$ and beam-corrected $\nu_{\rm peak}^{'}$ for both
blazars and radio galaxies, which supports the blazar sequence and
unification of blazars and radio galaxies (an alternative
relationship is the correlation between jet power and $\gamma$-ray
photon index).
\end{abstract}

\begin{keywords}
BL Lacertae objects: general -- galaxies: jets -- quasars: general
-- gamma-rays: galaxies
\end{keywords}

\section{Introduction}

Blazars are the most extreme and powerful active galactic nuclei
(AGN), pointing their jets in the direction of the observer (Urry \&
Padovani 1995; Ghisellini \& Tavecchio 2015). Based on the
equivalent width (EW) of the emission lines, the blazars are often
divided into two subclasses of BL Lacertae objects (BL Lacs; rest
frame EW$<5$ {\AA}) and flat spectrum radio quasars (FSRQs; Urry \&
Padovani 1995). Ghisellini et al. (2009, 2011) introduced a physical
distinction between the two classes of blazars, based on the
luminosity of the broad line region measured in Eddington units.
Giommi et al. (2012, 2013) suggested that blazars should be divided
into high and low ionization sources. Their broadband spectral
energy distributions (SEDs) are usually bimodal. The lower bump
peaks at IR-optical-UV band and the higher bump at GeV-TeV gamma-ray
band (Ghisellini et al. 1998; Abdo et al. 2010b).

Fossati et al. (1998) presented the blazar sequence: the peak
luminosity of the synchrotron component ($L_{\rm peak}$) and the
Compton dominance (the ratio of the peak of the Compton to the
synchrotron peak luminosity) are anti-correlation with the
synchrotron peak frequency $\nu_{\rm peak}$. Ghisellini et al.
(1998) modeled the broadband spectra of blazars, and suggested that
along the blazar sequence (BL Lacs $\rightarrow$ FSRQs), a stronger
radiative cooling causes a particle energy distribution with a break
at lower energies. So the synchrotron and Compton peaks shift to
lower frequencies. Various lines of evidence support the blazar
sequence (Georganopoulos et al. 2001; Cavaliere \& D'Elia 2002;
Maraschi \& Tavecchio 2003; Xie et al. 2007; Maraschi et al. 2008;
Ghisellini \& Tavecchio 2008; Ghisellini et al. 2009, 2010; Abdo et
al. 2010a; Chen \& Bai 2011; Sambruna et al. 2010; Finke et al.
2013; Chen 2014). But many contrary opinions have been presented
(Giommi, Menna \& Padovani 1999; Caccianiga \& Marcha 2004; Padovani
et al. 2003; Anton \& Browne 2005; Nieppola et al. 2006, 2008;
Padovani 2007; Giommi et al. 2012). The authors mainly focus on
sample selection effect and beaming effect. Ghisellini \& Tavecchio
(2008) revisited the blazar sequence and proposed that the SED of
blazar emission are linked to the mass of black hole and the
accretion rate. Finke (2013) studied a sample of blazars from 2LAC
(Ackermann et al. 2011) and found that a correlation exists between
Compton dominance and the peak frequency of the synchrotron
component for all blazars, including ones with unknown redshift. And
via constructing a model, Finke (2013) reproduced the trends of the
blazar sequence. Meyer et al. (2011) revisited the blazar sequence
and proposed the blazar envelope: FR I radio galaxies (FR Is) and
most BL Lacs belong to weak jet population while low synchrotron
peaking blazars and FR II radio galaxies (FR IIs) are strong jet
population.

Many evidences support that FR IIs are the parent population of
FSRQs, and FR Is are the parent population of BL Lacs (Urry \&
Padovani 1995). It is believed that the unification is mainly due to
relativistic beaming. Compared with blazars, radio galaxies is less
beaming with a larger jet angle to the line of sight. Fanaroff \&
Riley (1974) separated radio galaxies into two subclasses with
different morphology: the peak of low luminosity FR Is is close to
the nucleus, while high luminosity FR IIs have radio lobes with
prominent hot spots and bright outer edges. The distinction of
morphology also translates into a separation in radio power (the
fiducial luminosity $L_{\rm 178}\approx2\times10^{25}~{\rm
W~Hz^{-1}}$). Ledlow \& Owen (1994) found that the FR Is and FR IIs
break depends on radio and optical luminosity. In addition to host
galaxy magnitude and environments, FR Is and FR IIs differ in
optical emission lines (e.g. Zirbel \& Baum 1995). As in the case of
blazars, Ghisellini \& Celotti (2001) introduced that the division
between FR Is and FR IIs actually reflected a systematic difference
in accretion rate.

In this paper, after removing beaming effect, we revisited the
blazar sequence, and made use of the correlation between jet cavity
power and $\nu_{\rm peak}$ to study the unification of blazars and
radio galaxies. The paper is structured as follows: in Sect. 2, we
present the samples; the results are presented in Sect. 3;
discussions and conclusions are presented in Sect. 4. The
cosmological parameters $H_{\rm 0}=70~ {\rm km~s^{-1}~Mpc^{-1}}$,
$\Omega_{\rm m}=0.3$ and $\Omega_{\rm \Lambda}=0.7$ have been
adopted in this work.

\section{The samples}

Our sample was collected directly from sample of Nemmen et al.
(2012). But some blazars in Nemmen et al. (2012) were not clean
Fermi blazars (2LAC; Ackermann et al. 2011). We removed the
non-clean blazars. Generally, the SED was fitted by using a simple
third-degree polynomial function. However, many blazars were lack of
observed SED. Abdo et al. (2010b) have conducted a detailed
investigation of the broadband spectral properties of the
$\gamma$-ray selected blazars of the Fermi LAT Bright AGN Sample
(LBAS). They assembled high-quality and quasi-simultaneous SED for
48 LBAS blazars, and their results had been used to derive empirical
relationships that estimate the position of the two peaks from the
broadband colors (i.e. the radio to optical, $\alpha_{\rm ro}$, and
optical to X-ray, $\alpha_{\rm ox}$, spectral slopes) and from the
$\gamma$-ray spectral index. From
2LAC\footnote{http://www.asdc.asi.it/fermi2lac/}, we collected the
$\alpha_{\rm ro}$ and $\alpha_{\rm ox}$. Then from the empirical
relationships of Abdo et al. (2010b), we calculated the $\nu_{\rm
peak}$ and corrected redshift for the $\nu_{\rm peak}$. We excluded
the blazars without $\alpha_{\rm ox}$. At the same time, using
another empirical formula of Abdo et al. (2010b), we estimated the
synchrotron peak flux from 5 GHz flux and $\nu_{\rm peak}$. The 5
GHz flux is assembled from NASA/IPAC Extragalactic Database: NED.
When more than one flux was found, we took the most recent one. The
flux is K-corrected according to $S_{\rm \nu}=S^{\rm {obs}}_{\rm
\nu}(1+z)^{{\rm \alpha-1}}$, where $\alpha$ is the spectral index
and $\alpha=0.0$. The luminosity is calculated from the relation
$L_{\rm \nu}=4\pi{d_{\rm L}}^2S_{\rm \nu}$, and $d_{\rm L}$ is the
luminosity distance. From Nemmen et al. (2012), for Fermi blazars,
$f_{\rm b}$ is the beaming factor ($f_{\rm b}=1-{\rm
cos}(1/\Gamma$)), where $\Gamma$ is the bulk Lorentz factor of the
flow, since blazars obey $\theta_{\rm j}<1/\Gamma$ (Jorstad et al.
2005; Pushkarev et al. 2009). Pushkarev et al. (2009) calculated the
variability Lorentz factors $\Gamma_{\rm var}$ from long-term radio
observation. The bulk Lorentz factors of Nemmen et al. (2012) were
collected from the results of Pushkarev et al. (2009) (the number
$\sim$20\%). For blazars without $\Gamma_{\rm var}$ obtained from
Pushkarev et al. (2009), they used the power-law fit of $f_{\rm
b}\approx5\times10^{-4}(L^{\rm obs}_{\rm 49})^{-0.39\pm0.15}$ as an
estimator for $f_b$ (significant level at the 3.6$\sigma$; $L^{\rm
obs}_{\rm 49}$ is observation $\gamma$-ray luminosity). Moreover,
making use of the relation between cavity power and extended radio
300 MHz luminosity and assuming $P_{\rm jet}=P_{\rm cav}$, they
estimated the jet kinetic power. So the jet power is free of beaming
effect. The uncertainty of jet kinetic power is 0.7 dex. Our beaming
factor (or bulk Lorentz factor) and jet kinetic power of blazars are
obtained from Nemmen et al. (2012). For radio galaxies, we included
the complete radio galaxies sample of Meyer et al. (2011). The
$\nu_{\rm peak}$ and $L_{\rm peak}$ of Meyer et al. (2011) were
estimated by fitting non-simultaneous average SED. The jet power of
Meyer et al. (2011) was calculated from cavity power as same as
Nemmen et al. (2012). The relevant data for Fermi blazars were
listed in Table 1 and for radio galaxies in Table 3 of Meyer et al.
(2011).

In total, we have a sample containing 184 clean Fermi blazars (98
FSRQs and 86 BL Lacs) and 41 radio galaxies (24 FR Is and 17 FR
IIs).

\begin{figure}
\includegraphics[width=95mm, height=70mm]{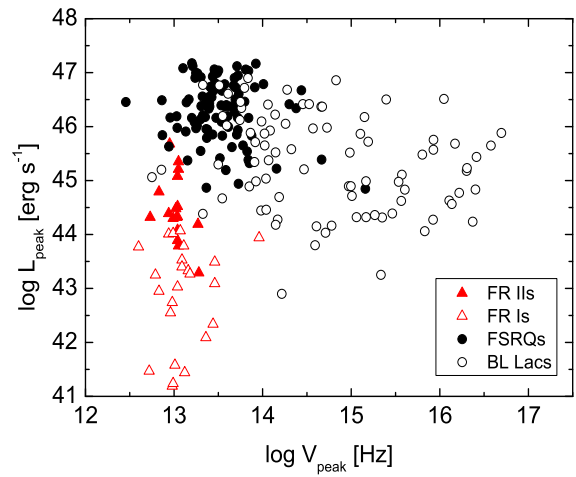}
\caption{synchrotron peak luminosity versus synchrotron peak
frequency. Fermi BL Lacs: black empty circles; Fermi FSRQs: black
filled circles; FR I radio galaxies: red empty triangles; FR II
radio galaxies: red filled triangles.} \label{figure 12}
\end{figure}

\begin{figure}
\includegraphics[width=160mm, height=120mm]{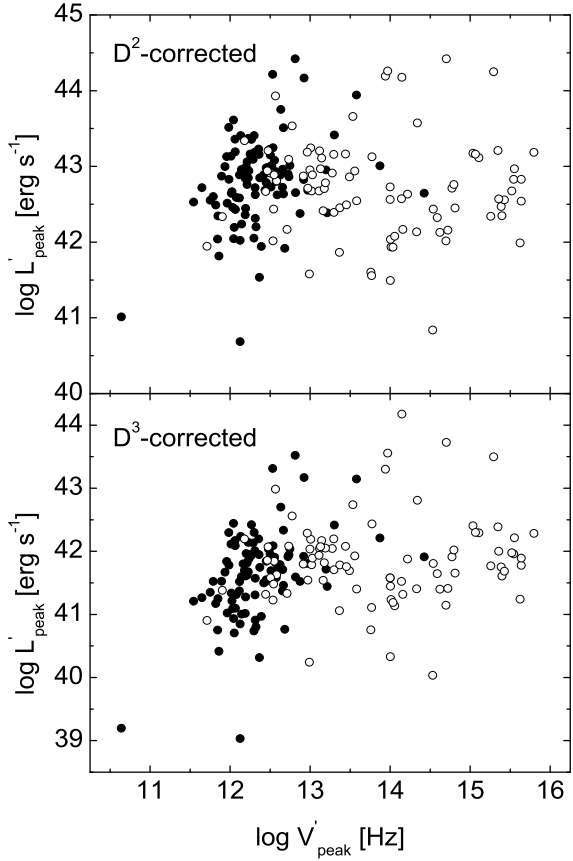}
\caption{Doppler-corrected synchrotron peak luminosity versus
Doppler-corrected synchrotron peak frequency for Fermi blazars. In
the top panel, the data-points are $D^2$-corrected and in the bottom
panel they are $D^3$-corrected. The meanings of different symbols
are as same as Fig. 1.} \label{figure 12}
\end{figure}

\section{The results}
\subsection{The blazar sequence}

The synchrotron peak luminosity versus synchrotron peak frequency is
shown in Fig. 1. A distinct ``L'' or ``V'' shape is not seen in this
figure. Blazars and radio galaxies are almost located in different
regions. The synchrotron peak luminosity $L_{\rm peak}$ of almost
all FSRQs (BL Lacs) is larger than $L_{\rm peak}$ of FR IIs (FR Is).
From Pearson correlation analysis, it is shown that for all blazars
sample, there is significant anti-correlation while not significant
for FSRQs; there is low anti-correlation for BL Lacs (see Table 2).
In Fig. 1, there is a BL Lac (J0630.9-2406: $\log \nu_{\rm
peak}=15.39$ Hz, $\log L_{\rm peak}=46.5~{\rm erg~s^{-1}}$) which is
found as high $L_{\rm peak}$ and high $\nu_{\rm peak}$ blazar in
Padovani, Giommi \& Rau (2012).

The Doppler corrections of synchrotron peak frequency and luminosity
scale as $\nu^{'}=\nu/ \delta$ and $L_{\rm peak}^{'}=L_{\rm peak}/
\delta^{\rm P}$ respectively, with $P=3+\alpha$ for a moving,
isotropic source and $P=2+\alpha$ for a continuous jet ($\delta$ is
Doppler factor; spectral index $\alpha=1$; Urry \& Padovani 1995 and
Nieppola et al. 2008). Since blazars obey $\theta_{\rm j}<1/\Gamma$,
especially for Fermi blazars (Jorstad et al. 2005; Pushkarev et al.
2009; Linford et al. 2011; Wu et al. 2014), we assume $\delta \sim
\Gamma$ ($\Gamma$ is estimated from radio observation or beaming
factor $f_{\rm b}$). From the above relations, Doppler-corrected
peak luminosity and frequency can be obtained ($D^2$-correction and
$D^3$-correction stand for $P=2+\alpha$ and $P=3+\alpha$
respectively). The Doppler-corrected synchrotron peak luminosity
versus Doppler-corrected synchrotron peak frequency is shown in Fig.
2. The results of Pearson correlation analysis show that for all
blazar and subclass, there are not significant correlations or are
significant positive correlations (both for $P=2+\alpha$ and
$P=3+\alpha$; see Table 2).

\begin{figure}
\includegraphics[width=95mm, height=70mm]{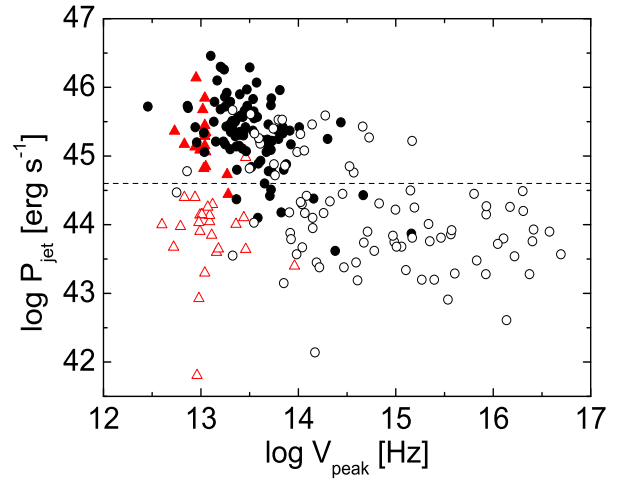}
\caption{Jet power versus non-beaming-corrected synchrotron peak
frequency. The uncertainty of jet power corresponds to 0.7 dex. The
discontinuous line corresponds to $\log P_{\rm jet}\sim44.6 ~{\rm
erg~s^{-1}}$. The meanings of different symbols are as same as Fig.
1.} \label{figure 12}
\end{figure}

\subsection{The unification of blazars and radio galaxies}

Sbarrato, Padovani \& Ghisellini (2014) have studied the question:
``how much do we have to beam the radio luminosity of the radio
galaxies to compare them with blazar?'' Sbarrato, Padovani \&
Ghisellini (2014) beamed the radio luminosity of the radio galaxies
as they were oriented as a blazar, who boosted the radio luminosity
of the radio galaxies by a factor:
\begin{equation}
~~~~~~~~~~~~~~~~~(\frac{\delta_{\rm BL}}{\delta_{\rm
RG}})^3=(\frac{1-\beta {\rm cos}\theta_{\rm RG}}{1-\beta {\rm
cos}\theta_{\rm BL}})^3,
\end{equation}
where $\delta$ is the Doppler-boosting factor with apparent
superluminal component velocities $\beta$. The Doppler-boosting
factor is expressed as $\delta=\frac{1}{\Gamma(1-\beta {\rm
cos}\theta)}$ with Lorentz factor $\Gamma=(1-\beta^2)^{-1/2}$ and
viewing angle $\theta$. Following the Sbarrato, Padovani \&
Ghisellini (2014), we boost the synchrotron peak frequency of radio
galaxies by a factor:
\begin{equation}
~~~~~~~~~~~~~~~~~\frac{\delta_{\rm BL}}{\delta_{\rm
RG}}=\frac{1-\beta {\rm cos}\theta_{\rm RG}}{1-\beta {\rm
cos}\theta_{\rm BL}}.
\end{equation}
We take average viewing angle $\theta_{\rm RG}\sim 40^\circ$ for
radio galaxies and $\theta_{\rm BL}\sim 3^\circ$ for blazars similar
to Sbarrato, Padovani \& Ghisellini (2014). If we assume jets of
blazar and radio galaxy are beamed with a Lorentz factor similar to
a common blazar, then the Lorentz factor $\Gamma\sim10$. If the jets
of radio galaxies and blazar are characterized by a rather small
Lorentz factor, then the Lorentz factor $\Gamma\sim3$. We also
estimate the factor $\frac{\delta_{\rm BL}}{\delta_{\rm RG}}$ from
assuming a average Lorentz factor $\Gamma\sim5$. The factors
$\frac{\delta_{\rm BL}}{\delta_{\rm RG}}$ are 4.7, 11.6 and 37
corresponding to $\Gamma\sim$3, 5 and 10 respectively. For radio
galaxies, we use $\nu_{\rm peak}^{'}=(\frac{\delta_{\rm
BL}}{\delta_{\rm RG}})\nu_{\rm peak}$ to beam their synchrotron peak
frequency (called beam-corrected synchrotron peak frequency). The
jet power versus non-beam-corrected synchrotron peak frequency and
beam-corrected synchrotron peak frequency can be seen in Fig. 3 and
Fig. 4 respectively. From Fig. 3 and Fig. 4, it is shown that when
the synchrotron peak frequency of radio galaxy is not
beam-corrected, the blazars have larger synchrotron peak frequency
than radio galaxies. Therefore most of blazars do not overlap with
radio galaxies. However after correcting beaming effect on peak
frequency of radio galaxies, we find that most of blazars overlap
with radio galaxies. From Table 2, we can get that for
non-beam-corrected peak frequency, there are significant
correlations between $P_{\rm jet}$ and $\nu_{\rm peak}$ for
blazars$+$radio galaxies, only blazars, only BL Lacs, only FSRQs;
for beam-corrected peak frequency of radio galaxies, there are
significant correlations between $P_{\rm jet}$ and $\nu_{\rm peak}$
for blazars$+$radio galaxies sample, and the correlations of
beam-corrected peak frequency are much stronger than the correlation
of non-beam-corrected peak frequency.

\begin{figure}
\includegraphics[width=95mm, height=70mm]{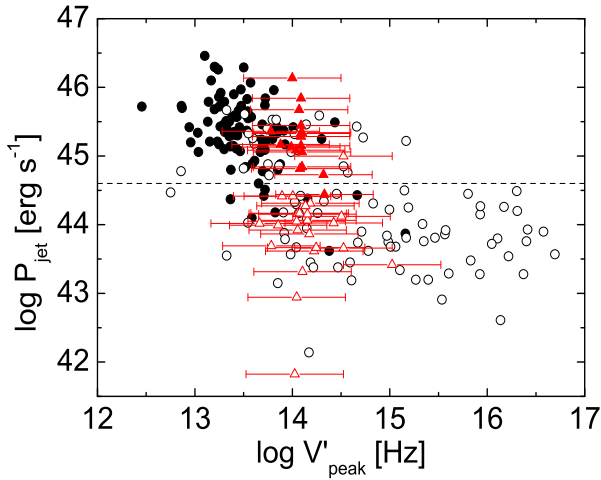}
\caption{Jet power versus beam-corrected synchrotron peak frequency.
The beam-corrected synchrotron peak frequency for radio galaxy is
estimated by assuming $\Gamma\sim5$ and viewing angle $\theta_{\rm
RG}\sim 40^\circ$ for radio galaxies and $\theta_{\rm BL}\sim
3^\circ$ for blazars. The error bar is beam-corrected synchrotron
peak frequency estimated by assuming Lorentz factors of 3 and 10.
The uncertainty of jet power corresponds to 0.7 dex. The
discontinuous line corresponds to $\log P_{\rm jet}\sim44.6 ~{\rm
erg~s^{-1}}$. The meanings of different symbols are as same as Fig.
1.} \label{figure 12}
\end{figure}

\section{Discussions and conclusions}

In order to study beaming effect on the blazar sequence and obtain
jet cavity power, we consider the sample of Nemmen et al. (2012)
which is large sample including beaming factor and jet cavity power
but not complete sample of 2LAC. Finke (2013) studied a 2LAC sample
(including blazars with unknown redshift), and used the empirical
equations of Abdo et al. (2010b) to estimate the $\nu_{\rm peak}$
and $L_{\rm peak}$. Via comparing our Fig. 1 with Fig. 2 of Finke
(2013), it is found that our Fig. 1 is consistent with Fig. 2 of
Finke (2013). The correlation analysis of blazars (with known
redshift) also supports that there is significant anti-correlation
between $\nu_{\rm peak}$ and $L_{\rm peak}$ which is consistent with
result of Finke (2013). In our Fig. 1, we also include radio
galaxies. From Meyer et al. (2011), an ``L''-shape in the $L_{\rm
peak}-\nu_{\rm peak}$ plot seems to have emerged which destroys the
blazar sequence. Also Meyer et al. (2011) presented the blazar
envelope (FR Is and BL Lacs belong to ``weak-jet'' sources; FR IIs
and FSRQs belong to ``strong-jet'' sources). However, combining
blazars and radio galaxies sample, we find that our results do not
have obvious the blazar envelope, especially for FR Is+BL Lacs
sample. The main reason is that there are still blazars in $10^{14}
{\rm Hz}<\nu_{\rm peak}<10^{15} {\rm Hz}$ interval for our sample.
Our $\nu_{\rm peak}$ and $L_{\rm peak}$ are estimated from the
empirical equations of Abdo et al. (2010b) who have pointed out that
their method assumed that the optical and X-ray fluxes are not
contaminated by thermal emission from the disk or accretion. In
blazars where thermal flux components are not negligible (this
should probably occur more frequently in low radio luminosity
sources) the method may lead to a significant overestimation of the
position of $\nu_{\rm peak}$. From Finke (2013), the following
factors can have effect on estimating $\nu_{\rm peak}$ and $L_{\rm
peak}$: the measurement error on determining $\nu_{\rm peak}$ and
$L_{\rm peak}$, the fitting function of SED and variability. Finke
(2013) have pointed out that the results in $L_{\rm peak}$ versus
$\nu_{\rm peak}$ plot are probably not accurate for individual
sources, whereas the overall trend still is present even for large
errors. From Equations (1), (3) and (4) of Abdo et al. (2010b), it
is in turn got that $\alpha_{\rm ro}\propto -logf_{\rm r}$, $\log
\nu_{\rm peak}\propto \log f_{\rm r}$, and $\log L_{\rm peak}\propto
\log f_{\rm r}$ ($f_{\rm r}$ is radio flux, $\alpha_{\rm ro}$ is
broadband spectral slopes between radio and optics). So even if the
correlation between $L_{\rm peak}$ and $\nu_{\rm peak}$ is dominated
due to both dependence on radio flux, the correlation should be
positive but not anti-correlation. In addition, when accounting for
the common dependence on radio flux (if any), we use Pearson partial
correlation analysis to analyze the correlation between $L_{\rm
peak}$ and $\nu_{\rm peak}$ for all blazars. The result shows that
the significant correlation still exist ($r=-0.47,
P=3\times10^{-11}$). In our sample, there is a BL Lac which is found
as high $L_{\rm peak}$ and high $\nu_{\rm peak}$ blazar in Padovani,
Giommi \& Rau (2012). Ghisellini et al. (2012) have found that the
four high $L_{\rm peak}$ and high $\nu_{peak}$ sources from
Padovani, Giommi \& Rau (2012) do not have excessively large
$\gamma$-ray luminosity or excessively small $\gamma$-ray photon
index. However, if the high $L_{\rm peak}$ and high $\nu_{\rm peak}$
blazars indeed exist, the blazars sequence based on $L_{\rm
peak}-\nu_{\rm peak}$ will be destroyed.

Another attention of the blazars sequence is beaming effect on
$L_{\rm peak}$ and $\nu_{\rm peak}$. Then the Doppler factor or bulk
Lorentz factor should be obtained. The variability Doppler factor
was derived from the associated variability brightness temperature
of total radio flux density flares with the intrinsic value of
brightness temperature, and the bulk Lorentz factor from apparent
super-luminal component velocities and variability Doppler factor
(Hovatta et al. 2009). For the blazars without without $\Gamma_{\rm
var}$ obtained from Pushkarev et al. (2009), Nemmen et al. (2012)
used $\gamma$-ray luminosity to estimate beaming factor with
3.6$\sigma$ significant level. For blazars with variability bulk
Lorentz factor available, the relative uncertainty in $\Gamma$ is
0.3 and the relative uncertainty in $\Gamma$ estimated from the
$L_{\rm \gamma}-f_{\rm b}$ relationship is 0.7 (Nemmen et al. 2012).
Furthermore, we assume $\Gamma\sim\delta$ which is reasonable
because $\theta_{\rm j}<1/\Gamma$ for blazars, especially for Fermi
blazars with a larger beaming factor and smaller $\theta_{\rm j}$
(Jorstad et al. 2005; Pushkarev et al. 2009; Linford et al. 2011; Wu
et al. 2014). It is noted that the above estimating $\delta$ is
probably not accurate for individual sources. From Fig. 2 and Table
2, we find that after being Doppler-corrected, for all blazar and
subclass, the correlations between $L_{\rm peak}$ and $\nu_{\rm
peak}$ are not significant or become significant positive
correlations, i.e. the anti-correction between $L_{\rm peak}$ and
$\nu_{\rm peak}$ disappears, which is consistent with result of
Nieppola et al. (2008).

Through studying the correlation between jet power and the
synchrotron peak frequency, it is found that there is significant
anti-correlation between $P_{\rm jet}$ and $\nu_{\rm peak}$ for
Fermi blazars, which supports the blazar sequence, i.e. a stronger
radiative cooling for higher jet power sources results in smaller
energies of the electrons emitting at the peaks. In addition, the
results also imply the BL Lacs - FSRQs divide. From Ghisellini et
al. (2009, 2011, 2015), Sbarrato et al. (2012), Sbarrato, Padovani
\& Ghisellini (2014) and Xiong \& Zhang (2014), it is got that
$P_{\rm jet}\sim L_{\rm disk}$ and the physical distinction between
BL Lacs and FSRQs is $L_{\rm BLR}/L_{\rm Edd}\sim 10^{-3}$. Based on
result of Xiong \& Zhang (2014), the average black hole mass of
Fermi blazars is $\sim10^{8.5}{\rm M_\odot}$. Then the jet power of
BL Lacs and FSRQs divide corresponds to $\log P_{\rm jet}\sim44.6
~{\rm erg~s^{-1}} (L_{\rm Edd}=1.3\times10^{38}(\rm
\frac{M}{M_\odot}){\rm erg~s^{-1}}, L_{\rm disk}\approx10L_{\rm
BLR}$, $M\sim10^{8.5}{\rm M_\odot}$). The Fig. 3 and Fig. 4 present
the divide line which also supports the physical distinction. The BL
Lacs with $\log P_{\rm jet}>44.6 ~{\rm erg~s^{-1}}$ (i.e. $L_{\rm
BLR}/L_{\rm Edd}> 10^{-3}$) can be considered as transition sources
(see Ghisellini et al. 2011). Moreover, the BL Lac which is found as
high $L_{\rm peak}$ and high $\nu_{\rm peak}$ blazar in Padovani,
Giommi \& Rau (2012) does not have high jet cavity power ($\log
P_{\rm jet}=43.2~{\rm erg~s^{-1}}$). A possible explanation is that
the high $L_{\rm peak}$ and high $\nu_{\rm peak}$ blazar is resulted
from beaming effect while the blazar virtually is low power source.
Ghisellini et al. (2009) have studied the $\gamma$-ray luminosity
versus $\gamma$-ray photon index (a proxy for inverse Compton peak
frequency), and proposed the Fermi blazars divide. We should note
that there is beaming effect on $\gamma$-ray luminosity while not on
jet power estimated from extended radio emission.

Next, we will discuss the unification of blazars and radio galaxies.
The Fig. 3 and Fig. 4 show that when the synchrotron peak frequency
of radio galaxy is not beam-corrected, most of blazars do not
overlap with radio galaxies (especially for FR Is and BL Lacs),
while after being beam-corrected peak frequency for radio galaxies,
most of blazars overlap with radio galaxies. Nieppola et al. (2008)
have found the dependence between Doppler factor and $\nu_{\rm
peak}$ for blazars. So the $\nu_{\rm peak}$ should be
Doppler-corrected or ruled out the influence of beaming effect.
However, it is not possible to derive the specific Lorentz
factor/Doppler factor for each source, especially for radio
galaxies. In this paper, we adopt the method of Sbarrato, Padovani
\& Ghisellini (2014). Making use of different Lorentz factor, we
analyze the correlations between jet power and beam-corrected peak
frequency for blazars$+$radio galaxies sample. From Table 2, we can
get that after being beam-corrected peak frequency of radio
galaxies, there are significant correlations between jet power and
peak frequency for blazars$+$radio galaxies sample, and the
correlations of beam-corrected peak frequency are much stronger than
the correlation of non-beam-corrected peak frequency. When
beaming-corrected peak frequency, the assuming Lorentz factors are
reasonable. For blazars, the $\Gamma\sim$3, 5 and 10 are possible
(see Sbarrato, Padovani \& Ghisellini (2014) for detail). Giovannini
et al. (2001) studied a complete sample of radio galaxies, and
estimated that relativistic jets with Lorentz factor in the range 3
- 10 are present in high and low power radio sources. Now, we
explain how to associate the correlations between jet power and
beam-corrected $\nu_{\rm peak}^{'}$ with the unification of blazars
and radio galaxies. Urry \& Padovani (1995) have explained that the
radio extended luminosity of FR IIs (FR Is) and FSRQs (BL Lacs)
should be comparable. Our jet cavity power is estimated from radio
extended luminosity. Then the jet power between FR IIs (FR Is) and
FSRQs (BL Lacs) should be comparable. In addition, as discussed in
the previous paragraph, the division between FSRQs and BL Lacs is
$\log P_{\rm jet}\sim 44.6 ~{\rm erg~s^{-1}}$ or $L_{\rm d}/L_{\rm
Edd}\sim10^{-2}$. According to the two-dimensional optical-radio
luminosity correlation from Owen \& Ledlow (1994), Ghisellini \&
Celotti (2001) found that the division between FR IIs and FR Is is
$P_{\rm jet}\sim0.015L_{\rm Edd}$. If the radio galaxies have
similar average black hole mass with blazars, then the division
between FR IIs and FR Is also is $\log P_{\rm jet}\sim 44.6 ~{\rm
erg~s^{-1}}$. From Fig. 3 and Fig. 4, it is seen that the line of
$\log P_{\rm jet}\sim 44.6 ~{\rm erg~s^{-1}}$ also can separate FR
Is from FR IIs. Abdo et al. (2010b) subdivided blazars based on the
$\nu_{\rm peak}$ (low/intermediate/high-synchrotron-peaked blazar).
Almost all FSRQs are low-synchrotron-peaked blazar and BL Lacs
include low/intermediate/high-synchrotron-peaked blazar. Based on
unified scheme, after being beam-corrected, the $\nu_{\rm peak}$
should be comparable between FR IIs (FR Is) and FSRQs (BL Lacs)
because the differences between FR IIs (FR Is) and FSRQs (BL Lacs)
mainly rely on jet angle to the line of sight. Therefore based on
above discussion, in $P_{\rm jet} - \nu_{\rm peak}^{'}$ plot,
blazars
\begin{figure}
\includegraphics[width=95mm, height=70mm]{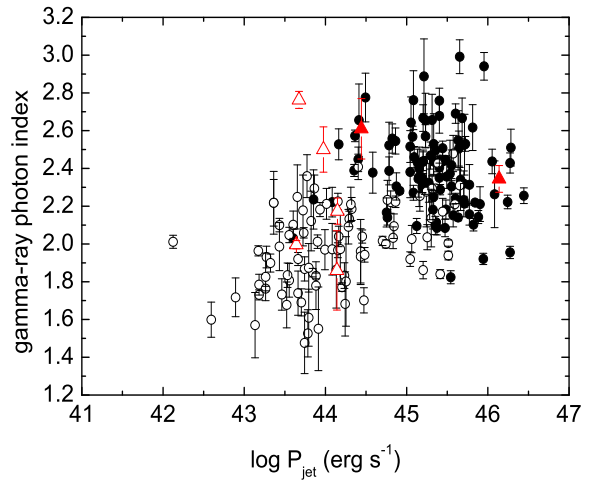}
\caption{$\gamma$-ray photon index versus jet power. The uncertainty
of jet power corresponds to 0.7 dex. The meanings of different
symbols are as same as Fig. 1.} \label{figure 12}
\end{figure}
and radio galaxies should have a similar correlation, i.e. the
blazars and radio galaxies can be unified from the correlation
between jet power and beam-corrected peak frequency. At present,
since only there are limited radio galaxies with precise Lorentz
factor estimated from observation, it is hard to accurately estimate
beam-corrected peak frequency for individual sources. In order to
test whether assuming Lorentz factors bring in remarkable effect on
the overall trend which results in spurious correlation, We also
analyze the correlation between jet power and $\gamma$-ray photon
index for Fermi blazar and radio galaxie because $\gamma$-ray photon
index is free of beaming effect and $L_{\rm \gamma}$ versus
$\gamma$-ray photon index has been used to support the blazar
sequence. If for blazars and radio galaxies, the correlations
between jet power and $\gamma$-ray photon index still exist and are
consist with the correlation between $P_{\rm jet}$ and
beam-corrected $\nu_{\rm peak}^{'}$, then it is clarified that our
assuming Lorentz factors do not bring in remarkable effect on the
overall trend. From Fig. 5 (FR Is: 3C 78, 3C 84, M87, Cen A, NGC
6251; FR IIs: 3C 380, 3C 111), the correlations between jet power
and $\gamma$-ray photon index still exist and support the blazar
sequence (higher jet power - softer photon index). In the
$\gamma$-ray photon index versus jet power plot, FR IIs overlap with
FSRQs and most of FR Is with BL Lacs, which support the unification
of blazars and radio galaxies, similar to the correlation between
$P_{\rm jet}$ and beam-corrected $\nu_{\rm peak}^{'}$. Therefore,
the assuming Lorentz factors do not bring in remarkable effect on
the overall trend.

\section*{Acknowledgments}

We sincerely thank anonymous referee for valuable comments and
suggestions. This work is financially supported by the National
Nature Science Foundation of China (11163007, U1231203, 11063004,
11133006 and 11361140347) and the Strategic Priority Research
Program ``The emergence of Cosmological Structures'' of the Chinese
Academy of Sciences (grant No. XDB09000000). This research has made
use of the NASA/IPAC Extragalactic Database (NED), that is operated
by Jet Propulsion Laboratory, California Institute of Technology,
under contract with the National Aeronautics and Space
Administration.

\begin{table*}
 \centering
  \caption{The Fermi blazar sample.}
  \begin{tabular}{@{}llrrrlrl@{}}
  \hline\hline
   2FGL name   &   Type    &   $Z^{\rm (a)}$   &   $\log P_{\rm jet}^{\rm (b)}$    &   $\log f_{\rm b}^{\rm (c)}$    &   $F_{\rm R}^{\rm (d)}$   &   $\log \nu_{\rm peak}^{\rm (e)}$ &   $\log L_{\rm peak}^{\rm (e)}$   \\
 \hline
J0050.6-0929    &   BLL &   0.634   &   44.31   &   -2.08   &   1920    &   14.83   &   46.86   \\
J0108.6+0135    &   FSRQ    &   2.099   &   46.46   &   -3.21$\ast$ &   3400    &   13.10   &   47.08   \\
J0112.1+2245    &   BLL &   0.265   &   43.75   &   -2.27   &   327     &   14.98   &   45.52   \\
J0112.8+3208    &   FSRQ    &   0.603   &   45.21   &   -2.52   &   405     &   13.15   &   45.37   \\
J0116.0-1134    &   FSRQ    &   0.67    &   45.46   &   -2.43   &   1900    &   13.30   &   46.13   \\
J0120.4-2700    &   BLL &   0.559   &   45.06   &   -2.45   &   1035    &   13.99   &   46.10   \\
J0132.8-1654    &   FSRQ    &   1.02    &   45.22   &   -2.65   &   1584    &   13.73   &   46.60   \\
J0136.9+4751    &   FSRQ    &   0.859   &   44.78   &   -2.62$\ast$ &   3150    &   13.69   &   46.73   \\
J0137.6-2430    &   FSRQ    &   0.838   &   45.57   &   -2.45   &   1558    &   13.58   &   46.37   \\
J0141.5-0928    &   BLL &   0.733   &   45.08   &   -2.41   &   1200    &   14.06   &   46.41   \\
J0205.3-1657    &   FSRQ    &   1.739   &   45.42   &   -2.79   &   1400    &   13.46   &   46.79   \\
\hline
\end{tabular}
\begin{quote}
(a) redshift directly collected from NED. (b) jet power estimated by
cavity power in unit of ${\rm erg~s^{-1}}$. (c) beaming factor
$f_{\rm b}=1-{\rm cos}(1/\Gamma)$. values with a `` $\ast$ ''
represent that they are directly from radio observation. (d) total
radio 5 GHz flux in mJy. (e) rest-frame synchrotron peak frequency
and luminosity in Hz and ${\rm erg~s^{-1}}$.

(f) This table is published in its entirety in the electronic
edition. A portion is shown here for guidance.
\end{quote}
\end{table*}

\begin{table*}
 \centering
  \caption{The results of correlation analysis.}
  \begin{tabular}{@{}llrrl@{}}
  \hline\hline
  Sample & $L_{\rm peak}$ versus $\nu_{\rm peak}^{\rm (a)}$    &  ~~~~~~$L_{\rm peak}^{\rm D2}$ versus $\nu_{\rm peak}^{\rm D(b)}$  &   ~~~$L_{\rm peak}^{\rm D3}$ versus $\nu_{\rm peak}^{\rm D(c)}$  & $P_{\rm jet}$ versus $\nu_{\rm peak}^{\rm (e)}$\\
 \hline
& $r^{(f)}$ ~~~~~~~~~ $P^{(f)}$ & $r$  ~~~~~~~~ $P$~~~~~ & $r$  ~~~~~~~~~ $P$~~~~~~~&$r$  ~~~~~~~~~ $P$~~~~~~~~~\\
 \hline
 Blazars & -0.5 ~~~ $4.2\times 10^{-13}$ & 0.009 ~~~~~~~ 0.9~~~ & 0.26 ~~~ $3\times 10^{-4}$ & -0.69 ~~$3\times 10^{-27}$\\
 BL Lacs & -0.24 ~~ 0.02 & -0.03 ~~~~~~~ 0.8~~~ & 0.12 ~~~~~~ 0.26~~~ &-0.45 ~~$1.7\times 10^{-5}$\\
 FSRQs   & -0.18 ~~ 0.07 & ~~~~~~~~~~~~~~~~~~0.33 ~~~ $1\times 10^{-3}$& 0.46 ~ $1.5\times 10^{-6}$ &-0.52 ~~$4.8\times 10^{-8}$\\
\hline
Sample& $P_{\rm jet}$ versus $\nu_{\rm peak}^{\rm D(d)}$ ($\Gamma\sim3$)& $P_{\rm jet}$ versus $\nu_{\rm peak}^{\rm D(d)}$ ($\Gamma\sim5$)&$P_{\rm jet}$ versus $\nu_{\rm peak}^{\rm D(d)}$ ($\Gamma\sim10$) & $P_{\rm jet}$ versus $\nu_{\rm peak}^{\rm (e)}$\\
  \hline
Blazars+RG & -0.59 ~~~ $2.6\times 10^{-22}$ &~~~~~~~~~~~~~~~-0.62 ~~ $1.2\times 10^{-25}$ & ~~~~~~~~~~~~~~~-0.64 ~~ $2.1\times 10^{-27}$&-0.50 ~~$2.6\times 10^{-15}$ \\
 \hline
\end{tabular}
\begin{quote}
(a) Synchrotron peak luminosity versus synchrotron peak frequency.
(b) Doppler-corrected synchrotron peak luminosity versus
Doppler-corrected synchrotron peak frequency ($D^2$-corrected). (c)
Doppler-corrected synchrotron peak luminosity versus
Doppler-corrected synchrotron peak frequency ($D^3$-corrected). (d)
jet power versus beam-corrected synchrotron peak frequency from
assuming $\Gamma\sim$3, 5 and 10. (e) jet power versus
non-beam-corrected synchrotron peak frequency.

(f) The $r$ is the Pearson correlation coefficient; $P$ is the
chance probability.
\end{quote}
\end{table*}

\end{document}